\pdfoutput=1
\documentclass[11pt]{article}

\usepackage[]{ACL2023}

\usepackage{times}
\usepackage{latexsym}

\usepackage[T1]{fontenc}
\usepackage[utf8]{inputenc}

\usepackage{microtype}

\usepackage{inconsolata}

\usepackage{amsmath}
\usepackage{algorithm,algpseudocode}
\usepackage{mathtools}

\usepackage{amssymb}
\DeclareMathSymbol{\shortminus}{\mathbin}{AMSa}{"39}

\usepackage{booktabs}
\usepackage{pifont}
\newcommand{\cmark}{\ding{51}}%

\usepackage{cancel}
\newcommand{\Sref}[1]{\S\ref{#1}}
\newcommand{\Fref}[1]{Figure~\ref{#1}}
\newcommand{\Tref}[1]{Table~\ref{#1}}

\usepackage{multirow}

\newcommand\aspace{\hspace{1.2em}}
\usepackage{color, colortbl}
\definecolor{LightCyan}{rgb}{0.88,1,1}
\usepackage{adjustbox}
\usepackage{array}
\newcolumntype{R}[2]{%
    >{\adjustbox{angle=#1,lap=\width-(#2)}\bgroup}%
    l%
    <{\egroup}%
}
\newcommand*\rot{\multicolumn{1}{R{45}{1em}}}% no optional argument here, please!
\usepackage{enumitem}
\setlist[itemize]{leftmargin=*}

\title{ESPnet-ST-v2: Multipurpose Spoken Language Translation Toolkit}

\author{Brian Yan*$^1$ \aspace Jiatong Shi*$^1$ \aspace Yun Tang$^2$ \aspace Hirofumi Inaguma$^2$ \\ 
\textbf{Yifan Peng}$^1$ \aspace \textbf{Siddharth Dalmia}$^1$ \aspace \textbf{Peter Pol\'ak}$^3$ \aspace \textbf{Patrick Fernandes}$^1$ \\
\textbf{Dan Berrebbi}$^1$ \aspace \textbf{Tomoki Hayashi}$^4$ \aspace \textbf{Xiaohui Zhang}$^2$ \aspace \textbf{Zhaoheng Ni}$^2$ \\ 
\textbf{Moto Hira}$^2$ \aspace \textbf{Soumi Maiti}$^1$ \aspace \textbf{Juan Pino}$^2$ \aspace \textbf{Shinji Watanabe}$^{1,5}$\\
$^1$Carnegie Mellon University \aspace $^2$Meta AI \\ $^3$Charles University \aspace $^4$Nagoya University \aspace $^5$Johns Hopkins University\\ 
\texttt{\{byan, jiatongs\}@cs.cmu.edu}
}

\begin{document}
\maketitle
\begin{abstract}
ESPnet-ST-v2 is a revamp of the open-source ESPnet-ST toolkit necessitated by the broadening interests of the spoken language translation community.
ESPnet-ST-v2 supports 1) offline speech-to-text translation (ST), 2) simultaneous speech-to-text translation (SST), and 3) offline speech-to-speech translation (S2ST) -- each task is supported with a wide variety of approaches, differentiating ESPnet-ST-v2 from other open source spoken language translation toolkits.
This toolkit offers state-of-the-art architectures such as transducers, hybrid CTC/attention, multi-decoders with searchable intermediates, time-synchronous blockwise CTC/attention, Translatotron models, and direct discrete unit models.
In this paper, we describe the overall design, example models for each task, and performance benchmarking behind ESPnet-ST-v2, which is publicly available at \url{https://github.com/espnet/espnet}.\footnote{Please see our documentation for \href{https://github.com/espnet/espnet/tree/master/egs2/TEMPLATE/st1}{ST/SST} and \href{https://github.com/espnet/espnet/tree/master/egs2/TEMPLATE/s2st1}{S2ST} to get started. Example models and tutorials are provided.}
\end{abstract}

\section{Introduction}

The objective of this project is to contribute to the diversity of the open-source spoken language translation ecosystem.
Toward this, we launched this ESPnet-ST-v2 update in collaboration with researchers working on Fairseq \cite{ott2019fairseq} and TorchAudio \cite{yang2021torchaudio}.
This project focuses on: offline speech-to-text (ST), simultaneous speech-to-text (SST), and offline speech-to-speech (S2ST).
These three spoken language translation tasks have drawn significant interest, as evidenced by rising IWSLT\footnote{\href{https://iwslt.org}{International Workshop on Spoken Language Translation}} shared task participation.

The ST task can be considered a base form of spoken language translation. %which is modified or extended in SST or S2ST.
Early approaches to ST stemmed from coupling statistical automatic speech recognition (ASR) \cite{huang2014historical} and text-to-text translation (MT) \cite{al1999statistical}, and this type of cascaded approach is still common in the neural network era \cite{bentivogli2021cascade, zhang2022ustc}.
End-to-end differentiable (E2E) approaches have recently emerged as an alternative offering greater simplicity and superior performance in some cases \cite{inaguma-etal-2021-espnet}; however, E2E approaches still benefit from techniques originating from ASR and MT \cite{gaido-etal-2021-ctc, inaguma2021source}.

SST modifies ST by imposing an additional streaming requirement, where systems are expected to produce textual outputs while incrementally ingesting speech input.
Both the aforementioned cascaded and end-to-end approaches to ST have been adapted for SST \cite{ma2020simulmt, iranzo2021streaming, chen2021direct}, although the more direct nature of the latter may be advantageous for latency-sensitive applications.
On the other hand, S2ST extends ST by producing target speech rather than target text.
Again, cascaded approaches of ST followed by text-to-speech (TTS) came first \cite{initial_speech2speech, black2002tongues} and E2E approaches followed \cite{jia2019direct, lee2022direct, jia2022translatotron, inaguma2022unity}, with the latter offering smaller footprints and greater potential to retain source speech characteristics.

Given the recent swell in E2E ST, SST, and S2ST research, we have revamped ESPnet-ST \cite{inaguma2020espnet} which previously only supported E2E ST. In particular, this work:
\begin{itemize}
    \item Implements ST, SST, and S2ST using common Pytorch-based modules, including encoders, decoders, loss functions, search algorithms, and self-supervised representations. 
    \item Builds a variety of example E2E models: attentional encoder-decoders, CTC/attention, multi-decoders with searchable intermediates, and transducers for ST. Blockwise attentional encoder-decoders, time-synchronous blockwise CTC/attention and blockwise transducers for SST. Spectral models (i.e. Translatotron) and discrete unit based models for S2ST.
    \item Benchmarks the ST, SST, and S2ST performance of ESPnet-ST-v2 against top IWSLT shared task systems and other prior works.
\end{itemize}
With this major update, ESPnet-ST-v2 keeps pace with the interests of the community and offers a variety of unique features, making it a valuable complement to Fairseq \cite{wang2020fairseq}, NeurST \cite{zhao2021neurst}, and other spoken language translation toolkits.

\section{Related Works}

ESPnet-ST-v2 follows a long line of open-source speech processing toolkits which can support spoken language translation \cite{zenkel2018open, shen2019lingvo, kuchaiev2019nemo, hayashi2020espnet, wang2020fairseq, zhao2021neurst}. 

In \Tref{tab:features} we compare ESPnet-ST-v2 to Fairseq \cite{wang2020fairseq} and NeurST \cite{zhao2021neurst}, two toolkits which also cover multiple types of spoken language translation. 
Fairseq and NeurST offer cascaded and E2E approaches to ST and SST (some of which are not offered by ESPnet-ST-v2). 
Meanwhile, ESPnet-ST-v2 focuses on E2E approaches and offers multiple unique core architectures not covered by the other toolkits.
For S2ST, Fairseq and ESPnet-ST-v2 both offer a range of approaches.
All told, ESPnet-ST-v2 offers the greatest variety across ST, SST, and S2ST -- however, we view these toolkits as complementary.
The following section elaborates on the unique features of ESPnet-ST-v2.

\begin{table}[t]
  \centering
    \resizebox {\linewidth} {!} {
\begin{tabular}{llllll}
\toprule
\textsc{Features} & \rot{\textsc{\textbf{ESPnet-ST-v2}}}  & \rot{\textsc{ESPnet-ST-v1}} & \rot{\textsc{Fairseq-S2T}} & \rot{\textsc{NeurST}} & \\

\midrule
\rowcolor{LightCyan}
\textbf{Offline ST} & \cmark & \cmark & \cmark & \cmark \\
End-to-End Architecture(s) & \cmark & \cmark & \cmark & \cmark \\
\hspace{2mm} Attentional Enc-Dec  & \cmark & \cmark & \cmark & \cmark \\
\hspace{2mm} CTC/Attention & \cmark & - & - & - \\
\hspace{2mm} Transducer & \cmark & - & - & -  \\
\hspace{2mm} Hierarchical Encoders & \cmark & - & - & - \\
\hspace{2mm} Multi-Decoder & \cmark & - & - & - \\
Cascaded Architectures & \cmark & \cmark & \cmark & \cmark \\
Speech SSL Representations & \cmark$^1$ & - & \cmark & - \\
Speech \& Text Pre-training & \cmark & \cmark & \cmark & \cmark \\
Joint Speech/Text Pre-training & - & - & \cmark & - \\
\midrule
\rowcolor{LightCyan}
\textbf{Simultaneous ST} & \cmark & - & \cmark & \cmark$^3$ \\
End-to-End Architecture(s) & \cmark & - & \cmark & -  \\
\hspace{2mm} Contextual Block Encoders & \cmark & - & - & - \\
\hspace{2mm} Blockwise Attn Enc-Dec & \cmark & - & - & - \\
\hspace{2mm} Blockwise CTC/Attention & \cmark & - & - & - \\
\hspace{2mm} Blockwise Transducer & \cmark & - & - & - \\
\hspace{2mm} Wait-K Attn Enc-Dec & - & - & \cmark & - \\
\hspace{2mm} Monotonic Attn Enc-Dec & - & - & \cmark & - \\
Cascaded Architectures & - & - & \cmark & \cmark$^3$ \\
\midrule
\rowcolor{LightCyan}
\textbf{Offline S2ST} & \cmark & - & \cmark & - \\
End-to-End Architecture(s) & \cmark & - & \cmark & - \\
\hspace{2mm} Spec Enc-Dec (Translatotron) & \cmark & - & \cmark & - \\
\hspace{2mm} Spec Multi-Dec (Translatotron 2) & \cmark & - & \cmark & - \\
\hspace{2mm} Discrete Enc-Dec (Speech-to-Unit) & \cmark & - & \cmark & - \\
\hspace{2mm} Discrete Multi-Decoder (UnitY) & \cmark & - & \cmark & - \\
Speech SSL Representations & \cmark$^1$ & - & \cmark & - \\
Neural Vocoder Support & \cmark$^2$ & \cmark & \cmark & - \\
\bottomrule
\end{tabular}
}

    \vspace{-3mm}
    \caption{Key features of ESPnet-ST-v2 compared to ESPnet-ST-v1 \cite{inaguma2020espnet}, Fairseq \cite{wang2020fairseq}, and NeurST \cite{zhao2021neurst}. Comparison intends to highlight unique features of ESPnet-ST-v2 and not to comprehensively review all toolkits. $^1$Supports S3PRL \cite{yang2021superb}. $^2$Supports both spectral and discrete. $^3$Only supports text-to-text.}
    \label{tab:features}
\end{table}

\section{ESPnet-ST-v2}

\begin{figure}
  \centering
  \includegraphics[width=\linewidth]{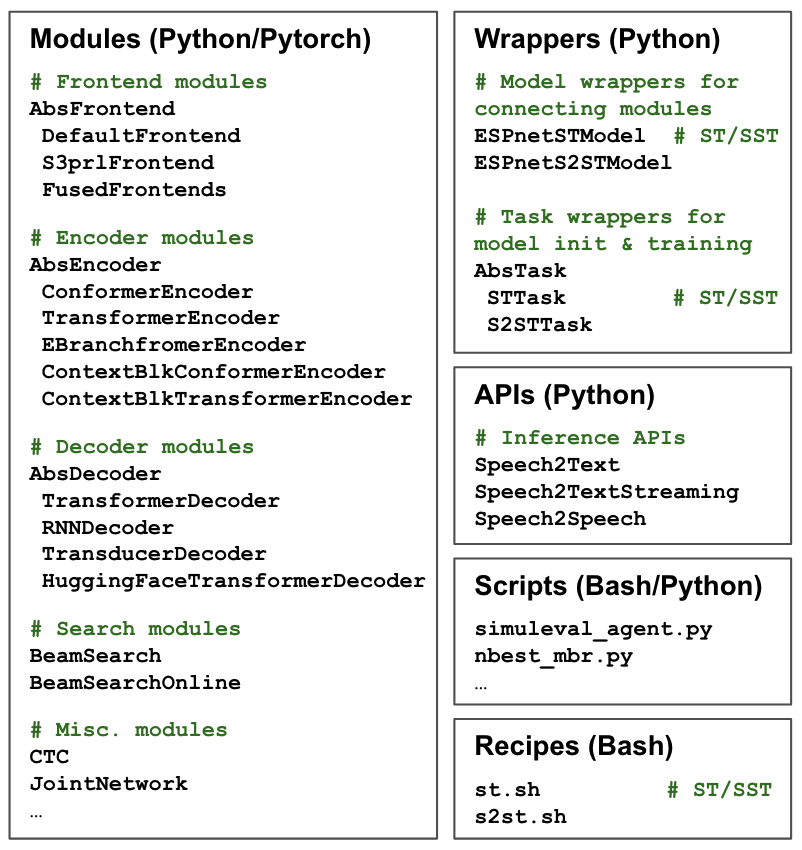}
  \caption{Software architecture of ESPnet-ST-v2.}
  \label{fig:design}
\end{figure}

In this section, we first describe the overall design and then introduce a few key features. 

\subsection{Modular Design}
\Fref{fig:design} illustrates the software architecture of ESPnet-ST-v2.
This modular design is an improvement over the ESPnet-ST-v1 where monolithic model and task definitions made it more difficult to extend and modify the toolkit.
We also designed ESPnet-ST-v2 such that modules developed for adjacent tasks (e.g. ASR, TTS, MT) can also be readily used for spoken language translation. 

In ESPnet-ST-v2 major neural network modules, such as frontends, encoders, decoders, search, and loss functions, inherit from common abstract classes making them easy to interchange.
These modules, which are detailed further in the next subsection, are used as building blocks in wrapper classes which are used to construct model architectures.
Then the fully constructed models are fed to task wrappers which prepare data loaders, initialize models, and handle training/validation.
For inference, pythonic APIs invoke search algorithms over the trained models and direct outputs to scoring scripts. 
For instance, the third-party SimulEval tool for evaluating SST latency \cite{simuleval2020} is integrated via this API layer.
We are also integrating with TorchAudio \cite{yang2021torchaudio} in the same manner.
Finally, recipe scripts define experimental pipelines from data preparation to evaluation.

\subsection{Key Features}

Each of the following modeling components feature a variety of interchangeable approaches.

\paragraph{Frontends \& Targets} Spectral features (e.g. FBANK) and features extracted from speech self-supervised learning (SSL) representations are supported, as well as fusions over multiple features \cite{berrebbi22_interspeech}.
For speech SSL features, ESPnet-ST-v2 integrates with the S3PRL toolkit \cite{yang2021superb}.
These speech SSL representations are also used to generate discrete targets for S2ST \cite{lee2022direct}.

\paragraph{Encoder Architectures} Conformer \cite{gulati2020conformer, guo2021recent}, Branchformer \cite{peng2022branchformer}, EBranchformer \cite{kim2023branchformer}, and Transformer \cite{vaswani2017attention, karita2019comparative} encoder architectures are supported for ST and S2ST. 
For SST, a blockwise scheme is adopted following \cite{tsunoo2021streaming, deng22b_interspeech} to form contextual block Conformer and Transformer encoders. 
Intermediate CTC \cite{lee2021intermediate} and Hierachical CTC \cite{sanabria2018hierarchical} encoding are also supported; these techniques have been shown to stabilize deep encoder optimization \cite{lee2021intermediate} and improve representations for sequence tasks involving source-to-target re-ordering \cite{yan2022ctc}.
% Finally, SSL encoders via S3PRL \cite{yang2021superb} can be used directly or fine-tuned.

\paragraph{Decoder Architectures} Attentional Transformer and recurrent neural network decoders are supported \cite{karita2019comparative}.
Multi-decoder schemes which allow for E2E differentiable decoder cascades via searchable hidden intermediates \cite{dalmia-etal-2021-searchable}, are also supported; this technique has been shown to improve sequence modeling for tasks which naturally decompose into sub-tasks.
Finally, large language model decoders (e.g. mBART \cite{liu2020multilingual}) can be adopted through an integration with HuggingFace \cite{wolf-etal-2020-transformers}.

\paragraph{Loss Functions} Cross-entropy (for attentional decoders), CTC, and Transducer are supported for ST and SST. Multi-objective training with CTC/attention and CTC/transducer as well as multi-tasked training (e.g. ASR/MT/ST) is also supported. For S2ST, L1 and mean square error losses are also supported for spectral models.

\paragraph{Search Algorithms} For offline attentional decoder models, label-synchronous beam search is supported with optional CTC joint decoding for multi-objective models \cite{watanabe2017hybrid}. 
For offline Transducer models, the original Graves beam search \cite{Graves2012SequenceTW} as well as time-synchronous and alignment-synchronous beam search \cite{saon2020alignment} beam searches are supported.
For SST, both incremental decoding and non-incremental (allowing re-translation) decoding \cite{liu20s_interspeech} are supported, along with stable hypothesis detection methods \cite{polak2022cuni}.
Blockwise attentional decoder models use a label-synchronous beam search or time-synchronous beam search if a CTC branch is available.
Blockwise transducer models use time-synchronous beam search.

\paragraph{Synthesis \& Post-processing}
For ST, Minimum Bayes Risk (MBR) ensembling \cite{fernandes2022quality} is supported for leveraging quality-metrics (e.g. BLEU) to compare and rank n-best outputs from one or more models.
For S2ST, neural vocoders are supported for both spectral and discrete inputs \cite{hayashi2020espnet, hayashi2021espnet2}.

\section{Example Models}
\label{sec:examples}

In this section, we introduce example models which are pre-built in ESPnet-ST-v2 using the neural network components described in the previous section. 
These examples include state-of-the-art core architectures, as evidenced by prior studies and our performance benchmarking (presented in \Sref{sec:exp}).

\subsection{ST Models}
\label{sec:st_models}

\begin{figure}[t]
  \centering
  \includegraphics[width=\linewidth]{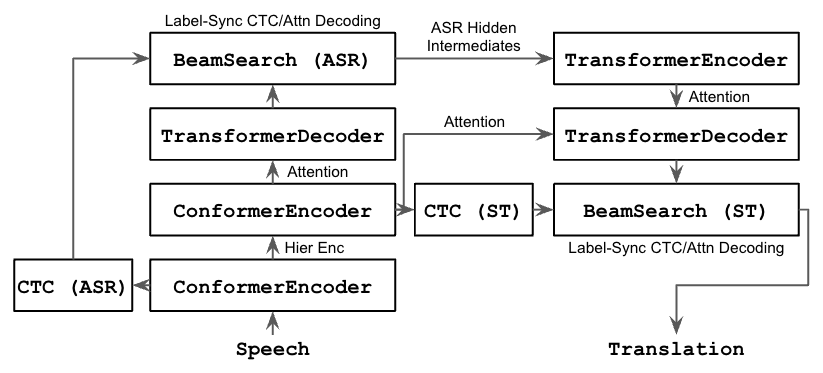}

  \caption{Multi-Decoder CTC/Attention for ST.}
  \label{fig:example-a}
\end{figure}

\paragraph{CTC/Attention (CA)} Following \citet{yan2022ctc}, we use Conformer encoders with hierarchical CTC encoding and Transformer decoders. 
The hierachical CTC encoding, which aligns the first $N$ layers of the encoder towards ASR targets and the last $M$ layers towards ST targets, regularizes the final encoder representations to be monotonic with respect to the target.
CTC/attention models are jointly decoded using either label-synchronous (wherein the attention branch is primary) or time-synchronous (wherein the CTC branch is primary) beam search.
For offline tasks, label-synchrony has shown greater performance \cite{yan2022ctc}.

\paragraph{Multi-Decoder CTC/Attention (MCA)} As shown in \Fref{fig:example-a}, the Multi-decoder decomposes ST into two sub-tasks, logically corresponding to ASR and MT encoder-decoder models, while maintaining E2E differentiability \cite{dalmia-etal-2021-searchable}.
This Multi-decoder scheme is also combined with the CTC/attention scheme described in the blurb above, following \citet{yan2022cmu}. We use Conformer encoders with hierarchical CTC for encoding speech and Transformer encoders for encoding intermediate ASR text. 
We use Transformer decoders for both ASR and ST.
During inference, the ASR stage is decoded first and then the final MT/ST stage is decoded; both stages use label-synchronous joint CTC/attention beam search.

\subsection{SST Models}

\begin{figure}[t]
  \centering
  \includegraphics[width=.8\linewidth]{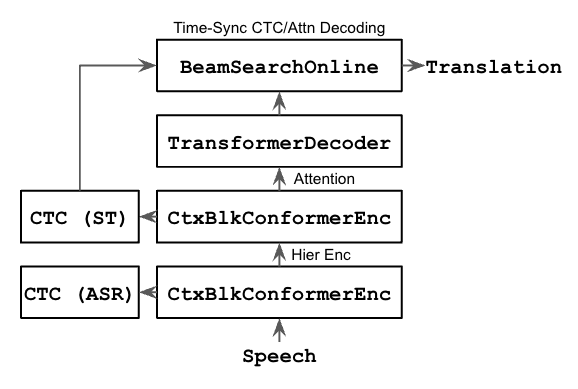}

  \caption{Time-Sync Blockwise CTC/Attn for SST.}
  \label{fig:example-b}
\end{figure}

\paragraph{Time-Synchronous Blockwise CTC/Attention (TBCA)} As shown in \Fref{fig:example-b}, we adapt the aforementioned CTC/attention model for ST (\Sref{sec:st_models}) to SST by replacing the Conformer encoder with a contextual block Conformer \cite{tsunoo2021streaming}.
During inference, we initially followed \citet{deng22b_interspeech} and used the label-synchronous CTC/attention beam search originally proposed for ASR by \citet{tsunoo2021streaming}.
However, we found that label-synchrony results in overly conservative boundary block detection for SST.
Therefore we opt instead for the time-synchronous variant which relies on CTC's more robust end-detection \cite{yan2022ctc} to control boundary block detection; this change reduces latency without sacrificing quality.
To perform incremental decoding without re-translation (as expected by SimulEval), hypotheses are pruned after processing all of the time steps for each encoder block.

\paragraph{Blockwise Transducer (BT)} As demonstrated by \citet{xue22d_interspeech}, Transducers can be effectively applied to SST despite the monotonic nature of their underlying alignment model.
We build Transducers for SST using contextual block Conformer encoders and unidirectional LSTM decoders.
We found that the aforementioned hierarchical CTC encoding (\Sref{sec:st_models}) improves training stability and convergence rate.
During inference, we found that the time-synchronous algorithm described by \citet{saon2020alignment} outperformed the original Graves decoding \cite{Graves2012SequenceTW} and the later proposed alignment-synchronous algorithms \cite{saon2020alignment}. 
We also found that length normalization is required to avoid overly short outputs.
Incremental decoding is applied in the same manner as for TBCA.

\begin{table*}
  \centering
    \resizebox {.95\linewidth} {!} {
\begin{tabular}{llcccc}
\toprule

\textsc{Toolkit} & \textsc{Model Type} & \textsc{DE} & \textsc{ES} & \textsc{FR} & avg \\% & \textsc{IT} \\
\midrule
\multicolumn{2}{c}{\textsc{Offline Speech Translation (ST)}} & \multicolumn{4}{c}{\textsc{BLEU $\uparrow$}} \\
\cmidrule(lr){1-2}\cmidrule(lr){3-6}

NeurST \cite{zhao2021neurst} & Attentional Enc-Dec (AED) & 22.8 & 27.4 & 33.3 & 27.8 \\%& 22.9\\% & - & - \\
Fairseq \cite{wang2020fairseq} & Attentional Enc-Dec (AED) & 22.7 & 27.2 & 32.9 & 27.6 \\%& 22.7 \\%& - & - \\
ESPnet-ST-v1 \cite{inaguma2020espnet} & Attentional Enc-Dec (AED) & 22.9 & 28.0 & 32.8 & 27.9 \\%& 23.8 \\%& - & - \\
\textbf{ESPnet-ST-v2} (this work) & Multi-Decoder CTC/Attn (MCA) & \textbf{27.9} & \textbf{32.1} & \textbf{38.5} & \textbf{32.8}\\

\midrule

\multicolumn{2}{c}{\textsc{Simultaneous Speech Translation (SST)}} & \multicolumn{4}{c}{\textsc{BLEU $\uparrow$ / AL $\downarrow$}} \\
\cmidrule(lr){1-2}\cmidrule(lr){3-6}
Fairseq \cite{wang2020fairseq} & Wait-K Attentional Enc-Dec (WAED) & 18.6 / 6.8 & 22.9 / 6.9 & 28.5 / 6.7 & 23.3 / 6.8 \\%& 15.4 / 6.8 \\%& - & - \\
\textbf{ESPnet-ST-v2} (this work) & Time-Sync Blockwise CTC/Attn (TBCA) & \textbf{23.5} / \textbf{2.3} & \textbf{29.2} / \textbf{2.4} & \textbf{32.7} / \textbf{2.3} & \textbf{28.5} / \textbf{2.3}  \\

\midrule

\multicolumn{2}{c}{\textsc{Offline Speech-to-Speech Translation (S2ST)}} & \multicolumn{4}{c}{\textsc{ASR-BLEU $\uparrow$}} \\
\cmidrule(lr){1-2} \cmidrule(lr){3-6}
Fairseq \cite{inaguma2022unity} & Discrete Multi-Decoder (UnitY) & \textbf{25.5} & \textbf{32.3} & 30.9 & \textbf{29.6} \\  % multiligual + from scratch
\textbf{ESPnet-ST-v2} (this work) & Discrete Multi-Decoder (UnitY) & 23.7 & 32.0 & \textbf{33.1} & \textbf{29.6} \\

\bottomrule

\end{tabular}
}

    \caption{Overview of ESPnet-ST-v2's ST, SST, and S2ST performances compared to other open-source toolkits.
    Results are presented on MuST-C-v1 (English-to-X) for ST/SST and on CVSS-C (X-to-English) for S2ST.}
    \label{tab:summary}
\end{table*}

\subsection{S2ST Models}

\begin{figure}[t]
  \centering
  \includegraphics[width=.8\linewidth]{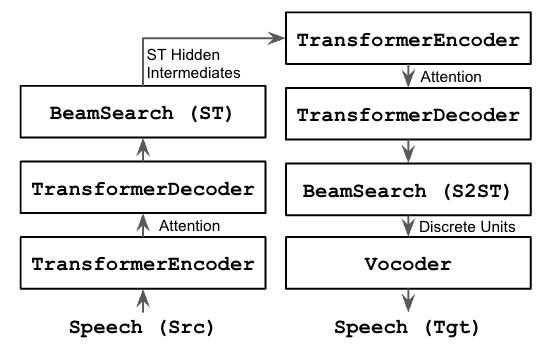}

  \caption{Discrete Multi-Decoder (UnitY) for S2ST.}
  \label{fig:example-c}
\end{figure}

\paragraph{Spectral Multi-Decoder (Translatotron 2)} Similar to the MCA model for ST (\Sref{sec:st_models}), the spectral Multi-decoder \cite{jia2022translatotron} decomposes S2ST into ST and TTS sub-tasks. 
The ST sub-task is modeled with an encoder-decoder network while the TTS sub-task is modeled with an auto-regressive synthesizer. 
The synthesizer attends over both the ST-encoder and ST-decoder hidden states. 
We use Transformers for the ST encoder-decoder and a Tacotron-style \cite{wang2017tacotron} decoder as the synthesizer. During inference, we first use beam search for the ST sub-task and then auto-regressively generate Mel-spectrograms. 
The final waveform speech is generated with a HiFi-GAN vocoder \cite{kong2020hifi}.

\paragraph{Discrete Multi-Decoder (UnitY)} The UnitY model \cite{inaguma2022unity} is similar to Translatotron 2, but critically predicts discrete units of speech SSL representations rather than spectral information in the final stage.
In other words, UnitY is Multi-decoder consisting of a ST sub-task followed by a text-to-unit (T2U) sub-task (see \Fref{fig:example-c}).
We use Transformer-based encoder-decoders for both sub-tasks.
During inference, the ST stage is first decoded and then followed by the T2U stage. 
Both stages use label synchronous beam search. 
The final speech is generated with a unit HiFi-GAN vocoder with Fastspeech-like duration prediction \cite{polyak21_interspeech, lee2022direct}, which is separately trained in the ParallelWaveGAN toolkit \cite{hayashi2020espnet, hayashi2021espnet2}.

\section{Performance Benchmarking}
\label{sec:exp}

In this section, we 1) compare open-source toolkits 2) compare our different example models and 3) compare our models with top IWSLT shared task systems and state-of-the-art prior works.

\subsection{Experimental Setup}

Please refer to \Sref{sec:repro} for reproducibility details.
The following is only a summary of our setup.

\paragraph{Data} We use MuST-C-v1 or MuST-C-v2 \cite{di-gangi-etal-2019-must} for ST/SST and CVSS-C for S2ST \cite{jia2022cvss}. 
For IWSLT comparisons, we combine MuST-C-v1, MuST-C-v2, and ST-TED \cite{niehues-etal-2018-iwslt} for ST/SST. 

\paragraph{Models} Unless otherwise indicated, we use a "base" setting for our models. Our base models have 40-80M trainable parameters across all tasks and are trained on a $\sim$400h of single language pair data from a single corpus. For ST/SST, we also use a "large" setting for benchmarking against IWSLT submissions. Our large models have 150-200M trainable parameters and are trained on $\sim$1000h of single language pair data from multiple corpora.

\paragraph{Scoring}
For ST/SST, we evaluate detokenized case-sensitive BLEU \cite{post-2018-call}. 
For SST, we additionally evaluate Average Lagging (AL) \cite{simuleval2020}.
For S2ST, we evaluate ASR-BLEU by transcribing the generated speech and then evaluating the BLEU of this transcription.

\begin{table}[t]
  \centering
    \resizebox {\linewidth} {!} {
\begin{tabular}{lcccc}
\toprule
\textsc{Model} & \textsc{HierEnc} & \textsc{BLEU}$\uparrow$ \\
\midrule
Attn Enc-Dec (AED) & - & 25.7 \\
Multi-Decoder Attn Enc-Dec (MAED) & - & 27.6 \\
CTC/Attention (CA) & \cmark & 28.6 \\
Multi-Decoder CTC/Attn (MCA) & \cmark & \textbf{28.8} \\
Transducer (T) & \cmark & 27.6 \\

\bottomrule
\end{tabular}
}

    \caption{Example ST models -- results on MuST-C-v2 En-De tst-COMMON.}
    \label{tab:offline}
\end{table}

\begin{table}[t]
  \centering
    \resizebox {.9\linewidth} {!} {
\begin{tabular}{lccccc}
\toprule
\textsc{Model} & \textsc{KD} & \textsc{BT} & \textsc{Ens} & \textsc{BLEU}$\uparrow$ \\
\midrule
\textbf{IWSLT'21 (Top 3 of 6)} \\
\texttt{1} Volctrans E2E$^\dag$ & \cmark & - & \cmark & \textbf{24.3} \\
\texttt{2} OPPO Cascade$^\dag$ & \cmark & \cmark & \cmark & 22.6 \\
\texttt{3} Volctrans Cascade$^\dag$ & \cmark & \cmark & \cmark & 22.2 \\
\midrule
\textbf{ESPnet-ST-v2} \\
\texttt{A} Base CA & - & - & - & 23.2 \\
\texttt{B} Base MCA & - & - & - & 23.6 \\
\texttt{C} Large CA & - & - & - & 24.3 \\
\texttt{D} Large MCA & - & - & - & \textbf{25.1} \\
\midrule
\texttt{E} MBR (\texttt{A+B+C+D}) & - & - & \cmark & \textbf{25.4} \\
\bottomrule
\end{tabular}
}

    \caption{Base and large CTC/attention (CA) and Multi-decoder CTC/attention (MCA) models compared to top IWSLT 2021 systems for the given segmentation tst2020 En-De test set. KD=Knowledge Distillation, BT=Back-Translation, Ens=Ensemble. $^\dag$Uses WMT MT data.}
    \label{tab:offline_iwslt}
\end{table}

\subsection{Results}

\paragraph{Toolkit Comparison} \Tref{tab:summary} summarizes ESPnet-ST-v2 performance, showing one best example model (\Sref{sec:examples}) for each task.
ESPnet-ST-v1, Fairseq, and NeurST models are also referenced for comparison.
On ST/SST, ESPnet-ST-v2 is 4-7 BLEU higher with 4.5 sec lower AL.\footnote{This comparison refers to the originally published results from the toolkit description papers. Note that subsequent works using these toolkits have improved the performance.}
On S2ST ESPnet-ST-v2 is on par with Fairseq.

\paragraph{ST} \Tref{tab:offline} shows a variety of approaches, of which the CTC/attention and Multi-decoder CTC/attention (MCA) models show the strongest performances.
In \Tref{tab:offline_iwslt}, we scale these two approaches by training on larger corpora and increasing model capacity -- \textit{our large MCA model outperforms the best IWSLT 2021 offline track submission on the 2020 test set with given segmentation.}

\begin{table}[t]
  \centering
    \resizebox {\linewidth} {!} {
\begin{tabular}{lc|ccc}
\toprule
\textsc{Model} & \textsc{BSz} & \textsc{BLEU}$\uparrow$/\textsc{AL}$\downarrow$ \\
\midrule
Blockwise Attn Enc-Dec (BAED) & 40 & 22.8 / 3.23 \\
Label-Sync Blockwise CTC/Attn (LBCA) & 40 & 24.4 / 3.23  \\
Time-Sync Blockwise CTC/Attn (TBCA) & 40 & \textbf{24.6} / \textbf{2.34}  \\
Blockwise Transducer (BT) & 40 & 22.9 / 2.37 \\
\midrule
Blockwise Attn Enc-Dec (BAED) & 20 & 21.0 / 2.77 \\
Label-Sync Blockwise CTC/Attn (LBCA) & 20 & \textbf{22.9} / 2.77 \\
Time-Sync Blockwise CTC/Attn (TBCA) & 20 & 22.8 / \textbf{1.63} \\
Blockwise Transducer (BT) & 20 & 20.9 / 1.71 \\
\bottomrule
\end{tabular}
}

    \caption{Example SST models -- results on MuST-C-v2 En-De tst-COMMON. BSz=Block Size.}
    \label{tab:streaming}
\end{table}

\begin{table}[t]
  \centering
    \resizebox {\linewidth} {!} {
\begin{tabular}{lcccccc}
\toprule
\textsc{Model} & \textsc{SSL} & \textsc{LLM} & \textsc{KD} & \textsc{BLEU}$\uparrow$ / \textsc{AL}$\downarrow$ \\
\midrule
\textbf{IWSLT'22  (Top 3 of 5)} \\
\texttt{1} CUNI-KIT E2E & \cmark & \cmark & - & \textbf{31.5} / \textbf{1.93} \\
\texttt{2} UPV Cascade$^\dag$ & - & - & - & 27.8 / 1.93 \\
\texttt{3} FBK E2E$^\dag$ & - & - & \cmark & 25.0 / 1.99 \\
\midrule
\textbf{ESPnet-ST-v2} \\
\texttt{A} Base TBCA & - & - & - & 24.7 / 1.93 \\
\texttt{B} Large TBCA & - & - & - & \textbf{26.6} / \textbf{1.93} \\
\bottomrule
\end{tabular}
}
    \caption{Base and large time-sync CTC/attention (TBCA) models compared to top IWSLT 2022 systems for the medium latency regime. Evaluated on En-De tst-COMMON-v2. SSL=Speech Self-Supervised Learning, LLM=Large Pre-trained Language Model, KD=Knowledge Distillation. $^\dag$Uses WMT MT data.}
    \label{tab:streaming_iwslt}

\end{table}

\paragraph{SST} \Tref{tab:streaming} shows a variety of approaches, of which the blockwise Transducer (BT) and time-synchronous blockwise CTC/attention (TBCA) models have the lowest AL. 
We choose to scale the TBCA to compare with IWSLT submissions due to its superior translation quality, but note that the BT has lower computational overhead due primarily to the lack of source-target computation; AL is non-computation aware.
In \Tref{tab:streaming_iwslt}, we fit the TBCA to the 2 second AL latency regime by selecting a blocksize of 32 and scale it with more data and model capacity -- \textit{our large TBCA model would have ranked 3rd out of 6 amongst IWSLT 2022 submissions without using any SSL / LLM representations or knowledge distillation.}

\begin{table}[t]
    \centering
    \resizebox {\linewidth} {!} {
\begin{tabular}{lcc}
\toprule
\textsc{Model} & \textsc{Type} &\textsc{ASR-BLEU$\uparrow$} \\
\midrule
\textbf{Prior Works}\\
\texttt{1} Translatotron \citep{jia2019direct} & Spectral & 14.4 \\
\texttt{2} Translatotron2 \citep{jia2022translatotron} & Spectral & 30.3 \\
\texttt{4} Speech-to-Unit \citep{lee2022direct} & Discrete & 30.8 \\
\texttt{5} UnitY \citep{inaguma2022unity} & Discrete & 32.3 \\
\midrule
\textbf{ESPnet-ST-v2} \\
\texttt{A} Attn Enc-Dec (Translatotron) & Spectral & 16.6\\
\texttt{B} Multi-Decoder (Translatotron2) & Spectral & 24.3 \\
\texttt{C} Attn Enc-Dec (Speech-to-Unit) & Discrete & 31.3\\
\texttt{D} Multi-Decoder (UnitY) & Discrete & \textbf{32.0} \\
\bottomrule
\end{tabular}
}
    \caption{Example S2ST models -- results on CVSS-C Es-En test set. Prior works shown for comparison. 
    }
    \label{tab:s2st_main}
\end{table}

\begin{table}[t!]
    \centering
    
\resizebox {0.8\linewidth} {!} {
\begin{tabular}{lcccc}
\toprule
\textsc{Frontend} & \textsc{Discrete Unit} & \textsc{ASR-BLEU$\uparrow$}\\

\midrule
FBANK & HuBERT & 14.8 \\
wav2vec2\dag  & HuBERT & 21.2 \\
HuBERT\dag  & HuBERT & 21.4 \\
mHuBERT  & HuBERT & 21.5 \\
WavLM\dag  & HuBERT & \textbf{22.8} \\
 \midrule
FBANK & WavLM & 15.0 \\
wav2vec2\dag  & WavLM & 21.6 \\
HuBERT\dag  & WavLM & 22.1 \\
mHuBERT & WavLM & 22.0 \\
WavLM\dag  & WavLM & \textbf{23.1} \\
\bottomrule
\end{tabular}
}
    \caption{Ablation on different types of SSL for the frontend and discrete unit portions of S2ST models. \dag Trained with large settings, others with base settings.}
    \label{tab:s2st_ssl}
\end{table}

\paragraph{S2ST} \Tref{tab:s2st_main} shows a variety of approaches compared to prior works with comparable architectures -- \textit{our S2ST models are generally on par with prior works which are considered state-of-the-art.}
In fact, all of our models slightly outperform their respective prior works except for Translatotron 2.
Further, in \Tref{tab:s2st_ssl} we ablate a range of SSL types for both the frontend and discrete units demonstrating the flexibility of our toolkit. 

\section{Conclusion}

We presented ESPnet-ST-v2 which now supports offline speech translation, simultaneous speech translation, and offline speech-to-speech translation.
ESPnet-ST-v2 will continue to grow to support the community's interests.
Future updates may include more new tasks, such as simultaneous speech-to-speech translation, and cross-toolkit integrations via TorchAudio.

\section*{Limitations}
The first set of limitations to be aware of are data-related.
Although prior works have shown the feasibility of building E2E systems without source language transcriptions \cite{lee-etal-2022-textless, chen2022speech, zhang2021uwspeech}, in this work we only investigate cases where triplet data (source speech, source transcript, target translation) is available for ST/SST and where quadruplet data (source speech, source transcript, target translation, target speech) is available for S2ST.

The second set of limitations to be aware of are evaluation-related.
For SST, we follow prior works \cite{simuleval2020, wang2020fairseq, anastasopoulos-etal-2022-findings} and evaluate AL which is a measure of how much the system outputs lags behind the amount of input read. 
Notably, this does not consider the actual computation time and only the input-to-output ratio.
For S2ST, we follow prior works \cite{jia2022translatotron, inaguma2022unity} and evaluate ASR-BLEU.
This evaluation is dependent on an ASR system, which is not standardized across prior works.
And further, our evaluation of S2ST outputs does not include naturalness.
Finally, in this work we have not conducted any human evaluation of translation outputs.

\section*{Acknowledgements}
Brian Yan and Shinji Watanabe are supported by the Human Language Technology Center of Excellence.
This work also used the Extreme Science and Engineering Discovery Environment (XSEDE) ~\cite{xsede}, which is supported by National Science Foundation grant number ACI-1548562; specifically, the Bridges system ~\cite{nystrom2015bridges}, as part of project cis210027p, which is supported by NSF award number ACI-1445606, at the Pittsburgh Supercomputing Center. 
This work also used GPUs donated by the NVIDIA Corporation.

\bibliography{anthology,custom}
\bibliographystyle{acl_natbib}

\appendix

\section{Appendix}
\label{sec:appendix}

\subsection{Reproducibility}
\label{sec:repro}

\begin{table*}[t]
  \centering
    \resizebox {\linewidth} {!} {
\begin{tabular}{lccccccccc}
\toprule
Model & Task & Encoder(s) & Decoder(s) & Frontend & Pre-Train Init & Multi-Obj & Src BPE & Tgt BPE & \# Params \\
\midrule
% ST
AED (\Tref{tab:offline}) & ST & 12 lyr, 4 head, 256 adim & (ASR) 6 lyr, 4 head & FBANK & ASR Enc/Dec & ASR & 4k & 4k & 60M \\
& & & (ST) 6 lyr, 4 head & &  \\
\midrule
MAED (\Tref{tab:offline}) & ST & (ASR) 12 lyr, 4 head, 256 adim & (ASR) 6 lyr, 4 head & FBANK & ASR Enc/Dec & ASR & 4k & 4k & 60M \\
& & (MT) 2 lyr, 4 head, 256 adim & (MT) 6 lyr, 4 head & &  \\
\midrule
CA (\Tref{tab:offline}) & ST & 18 lyr, 4 head, 256 adim & (ASR) 6 lyr, 4 head & FBANK & ASR Enc/Dec & ASR & 4k & 4k & 70M \\
& & & (ST) 6 lyr, 4 head & & \\
\midrule
MCA (\Tref{tab:offline}) & ST & (ASR) 18 lyr, 4 head, 256 adim & (ASR) 6 lyr, 4 head & FBANK & ASR Enc/Dec/CTC & ASR & 4k & 4k & 70M \\
& & (MT) 4 lyr, 4 head, 256 adim & (MT) 6 lyr, 4 head & & \\
\midrule
T (\Tref{tab:offline}) & ST & 18 lyr, 4 head, 256 adim & 1 lyr, 512 dim, 640 joint & FBANK & ASR Enc/Dec/CTC & ASR & 4k & 4k & 70M \\
& & & (ASR) 6 lyr, 4 head & & \\
\midrule
Large CA (\Tref{tab:offline_iwslt}) & ST & 18 lyr, 8 head, 512 adim & (ASR) 6 lyr, 4 head & HuBERT & ASR Enc/Dec/CTC & ASR & 8k & 16k & 210M \\
& & & (ST) 6 lyr, 4 head & & \\
\midrule
Large MCA (\Tref{tab:offline_iwslt}) & ST & (ASR) 18 lyr, 8 head, 512 adim & (ASR) 6 lyr, 8 head & HuBERT & ASR Enc/Dec/CTC & ASR & 8k & 8k & 210M \\
& & (MT) 4 lyr, 8 head, 512 adim & (MT) 6 lyr, 8 head & & \\

% SST
\midrule
BAED (\Tref{tab:streaming}) & SST & 18 lyr, 4 head, 256 adim & 6 lyr, 4 head & FBANK & ASR Enc lyr 1-12 & - & 4k & 4k & 70M \\
\midrule
LBCA (\Tref{tab:streaming}) & SST & 18 lyr, 4 head, 256 adim & 6 lyr, 4 head & FBANK & ASR Enc lyr 1-12 & - & 4k & 4k & 70M \\
\midrule
TBCA (\Tref{tab:streaming}) & SST & 18 lyr, 4 head, 256 adim & 6 lyr, 4 head & FBANK & ASR Enc lyr 1-12 & - & 4k & 4k & 70M \\
\midrule
BT (\Tref{tab:streaming}) & SST & 18 lyr, 4 head, 256 adim & 1 lyr, 4 head, 640 joint & FBANK & ASR Enc lyr 1-12 & - & 4k & 4k & 40M \\
\midrule
Large TBCA (\Tref{tab:streaming_iwslt}) & SST & 18 lyr, 8 head, 512 adim & 6 lyr, 8 head & FBANK & ASR Enc lyr 1-12 & - & 8k & 8k & 150M \\

% S2ST
\midrule
Translatotron (\Tref{tab:s2st_main}) & S2ST & 12 lyr, 4 head, 256 adim & 6 lyr, 1024 dim & FBANK & - & ASR, ST & 7k & 500 & 80M \\
\midrule
Translatotron2 (\Tref{tab:s2st_main}) & S2ST & 16 lyr, 4 head, 256 adim & (ST) 6 lyr, 4 head & FBANK & - & ASR, ST & 7k & 500 & 50M \\
& & & (TTS) 2 lyr, 1024 dim &  & & & & & \\
\midrule
Speech-to-Unit (\Tref{tab:s2st_main}) & S2ST & 12 lyr, 4 head, 512 adim & 6 lyr, 8 head & FBANK & - & ASR, ST & 7k & 500 & 40M \\
\midrule
UnitY (\Tref{tab:s2st_main}) & S2ST & (ST) 16 lyr, 4 head, 256 adim & (ST) 4 lyr, 4 head & FBANK & - & ASR, ST & 7k & 500 & 40M \\
& & (T2U) 2 lyr, 4 head, 256 adim & (T2U) 2 lyr, 8 head &  &  &  &  &  &  \\
\bottomrule
\end{tabular}
}
    \caption{ST, SST, and S2ST model hyperparameters. Parameter counts are rounded to the nearest 10 million.}
    \label{tab:models}
\end{table*}

\Tref{tab:models} shows the hyperparameters for the models presented in \Sref{sec:exp}.
All of our data preparation scripts are available in ESPnet: \url{https://github.com/espnet/espnet/tree/master/egs2}.

\end{document}